# Lateral oscillation and body compliance help snakes and snake robots stably traverse large, smooth obstacles

Qiyuan Fu, Sean W. Gart, Thomas W. Mitchel, Jin Seob Kim, Gregory S. Chirikjian, Chen Li*

Department of Mechanical Engineering, Johns Hopkins University

*corresponding author: [redacted]

## Synopsis

Snakes can move through almost any terrain. Similarly, snake robots hold the promise as a versatile platform to traverse complex environments like earthquake rubble. Unlike snake locomotion on flat surfaces which is inherently stable, when snakes traverse complex terrain by deforming their body out of plane, it becomes challenging to maintain stability. Here, we review our recent progress in understanding how snakes and snake robots traverse large, smooth obstacles like boulders and felled trees that lack "anchor points" for gripping or bracing. First, we discovered that the generalist variable kingsnake combines lateral oscillation and cantilevering. Regardless of step height and surface friction, the overall gait is preserved. Next, to quantify static stability of the snake, we developed a method to interpolate continuous body in three dimensions (both position and orientation) between discrete tracked markers. By analyzing the base of support using the interpolated continuous body 3-D kinematics, we discovered that the snake maintained perfect stability during traversal, even on the most challenging low friction, high step. Finally, we applied this gait to a snake robot and systematically tested its performance traversing large steps with variable heights to further understand stability principles. The robot rapidly and stably traversed steps nearly as high as a third of its body length. As step height increased, the robot rolled more frequently to the extent of flipping over, reducing traversal probability. The absence of such failure in the snake with a compliant body inspired us to add body compliance to the robot. With better surface contact, the compliant body robot suffered less roll instability and traversed high steps at higher probability, without sacrificing traversal speed. Our robot traversed large step-like obstacles more rapidly than most previous snake robots, approaching that of the animal. The combination of lateral oscillation and body compliance to form a large, reliable base of support may be useful for snakes and snake robots to traverse diverse 3-D environments with large, smooth obstacles.

## Introduction

Snakes can use their slender body with many degrees of freedom (Hoffstetter and Gasc 1969) to agilely move through almost any environment (Houssaye et al. 2013), such as deserts, grasslands, forests, and wetlands (Gray 1946; Jayne 1986; Marvi and Hu 2012). Snakes' ability to cope with complex 3-D environments has inspired the development of snake-like robots for critical tasks like search and rescue in earthquake rubbles, building inspection, and extraterrestrial exploration (Hirose 1993; Walker et al. 2016; Whitman et al. 2018). Despite research in arboreal locomotion (Byrnes and Jayne 2012; Hoefer and Jayne 2013; Jorgensen and Jayne 2017), burrowing (Sharpe et al. 2015), gliding (Socha 2002, 2011), and swimming (Graham and Lowell 1987; Munk 2008), we still understand relatively little about how snakes traverse complex 3-D terrain such as felled trees and boulders (Gart et al. 2019).

When snakes and snake robots use planar gaits to move on flat surfaces (Gray 1946; Jayne 1986; Marvi and Hu 2012), they are inherently stable (Dowling 1996; Hirose and Mori 2004). However, the more they bend the body out of plane, the more challenging it is to maintain stability (Hatton and Choset 2010; Byrnes and Jayne 2012; Toyoshima and Matsuno 2012; Hoefer and Jayne 2013; Marvi et al. 2014; Jorgensen and Jayne 2017). Arboreal snakes can grip branches and twigs, and desert snakes can brace against depressed sand, thereby using or creating "anchor points" for stability. Similarly, snake robots can use or create anchor points in complex environments like ladders, pipes, and desert dunes (Lipkin et al. 2007; Marvi et al. 2014; Takemori et al. 2018a). In contrast, it is less known how snakes or snake robots can stably traverse large, smooth obstacles lacking such anchor points, such as boulders and felled trees.

Here, we review our recent progress towards understanding this problem through interdisciplinary research. The key findings are: (1) the generalist kingsnake combines lateral oscillation and cantilevering to traverse large, smooth obstacles like steps (Gart et al. 2019); (2) continuous body 3-D reconstruction reveals that lateral oscillation creates large base of support, helping the animal to achieve perfect stability (Mitchel et al. 2020); and (3) robotic modeling reveals that body compliance improves terrain contact and helps maintain stability (Fu and Li 2020). Thanks to these bio-inspired strategies, our robot also achieves better traversal performance than most previous snake robots, approaching that of animals (Fu and Li 2020).

## Snakes use lateral oscillation and cantilevering to traverse large steps

To investigate the stability principle of snakes traversing large, smooth obstacles, we challenged the variable kingsnake [*Lampropeltis mexicana* (Garman 1883)], a generalist found in diverse rocky environments, to traverse a large step obstacle (Gart et al. 2019) (Fig. 1A). To test how the animal responded to terrain variation, we used two step heights, $H$ = 5 cm or 15% snout-vent length (SVL) and $H$ = 10 cm or 30% SVL and covered the steps with either high friction burlap or low friction paper. Note that for this challenging obstacle without available anchor points, the maximum traversable height of 30% SVL was not comparable to that of arboreal specialists bridging vertical gaps which are larger than 70% SVL (Byrnes and Jayne 2012). Three individuals were tested (SVL = 34.6 ± 0.4 cm, body length = 39.6 ± 0.4 cm, mean ± s.d.), each with 10 trials on each of the four step treatments, resulting in a total of 120 trials. Multiple markers were attached to the snake body. Their 3-D position and orientation were reconstructed using tracking from multiple camera views. See (Gart et al. 2019) for detail of animal experiments.

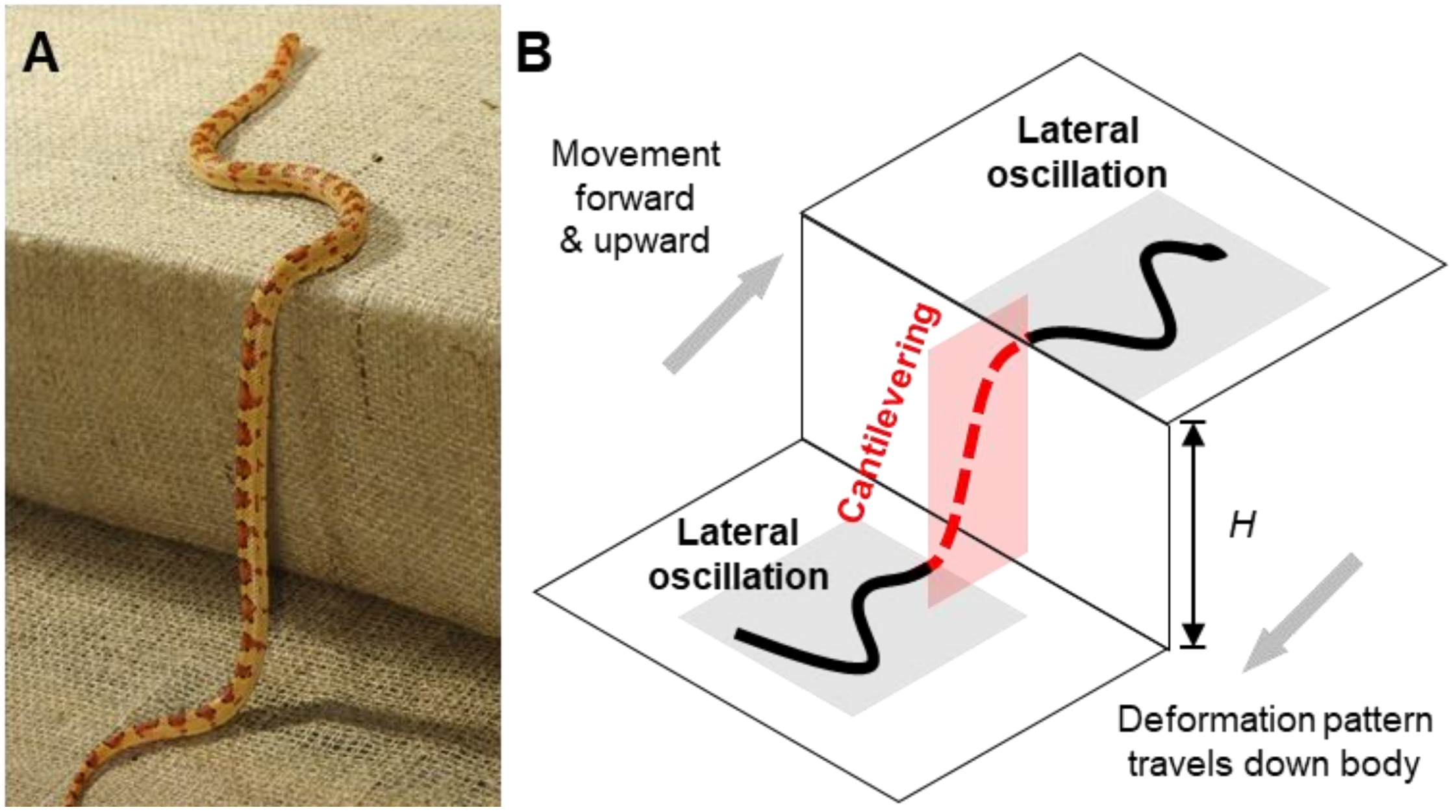

**Fig. 1. A kingsnake traversing a large step, using a partitioned gait combining body lateral oscillation and cantilevering. (A)** A representative snapshot of a kingsnake traversing a large high friction step (covered with burlap) as high as 30% snout-vent length (oblique view). **(B)** Partitioned gait template combining lateral oscillation and cantilevering.

We discovered that the snakes traversed large step obstacles by partitioning its body into three sections (Fig. 1A). Both the anterior and posterior body sections oscillated laterally on the horizontal surfaces above and below the step (Fig. 1B, black solid). To bridge the large height increase, the body section in between was suspended in the air. Initially, before the head reached the surface above (no anterior section yet), this section cantilevered in the air with one point of support. After the head reached the surface above, the middle body section was suspended between two points of support (Fig. 1B, red dashed). Whether it was cantilevering or suspended, the sagittal plane of this section was nearly vertical, and its shape was nearly constant throughout the entire traversal. We note that muscle activation may differ between a cantilevering and a suspended section (Jorgensen and Jayne 2017). However, given their similar kinematics, for simplicity, hereafter we refer to this middle body section as the cantilevering section. The overall partitioned gait pattern traveled down the body as the snake progressed forward and upward onto the step.

As step height and surface friction varied, the snake made fine adjustments of its partitioned gait. When step height increased, the snake devoted a longer body section to cantilevering and pitched it up more to accommodate the larger height. When surface friction decreased, the snake suffered large lateral and backward slip and moved more intermittently in response; its cantilevering body section was also closer to the step. However, despite these active adjustments, the overall partitioning the body into lateral oscillation and cantilevering was conserved.

**Stability advantage of lateral oscillation while cantilevering**

Many previous snake robots traversed large, smooth step-like obstacles using a follow-the-leader gait to simplify control and/or better utilize active propellers (Kurokawa et al. 2008; Birkenhofer 2010; Takaoka et al. 2011; Tanaka et al. 2018). With each body segment following the previous one, these robots often simply bend the body within a vertical (sagittal) plane to traverse (Fig. 2A). To be statically stable during cantilevering before the head reaches the surface above, careful feedback control is necessary to ensure that the center of mass always projects into the narrow base of support formed by the straight body in contact with the surface below (Fig. 2A). Otherwise, the robot can be easily tipped over by a lateral perturbation, such as slipping on low friction steps. As a result, these previous snake robots are often slow in traversing large step obstacles. In contrast, the snake's lateral oscillation during cantilevering may be key to its success, as it can help achieve a wide base of support to resist lateral perturbations (Fig. 2B).

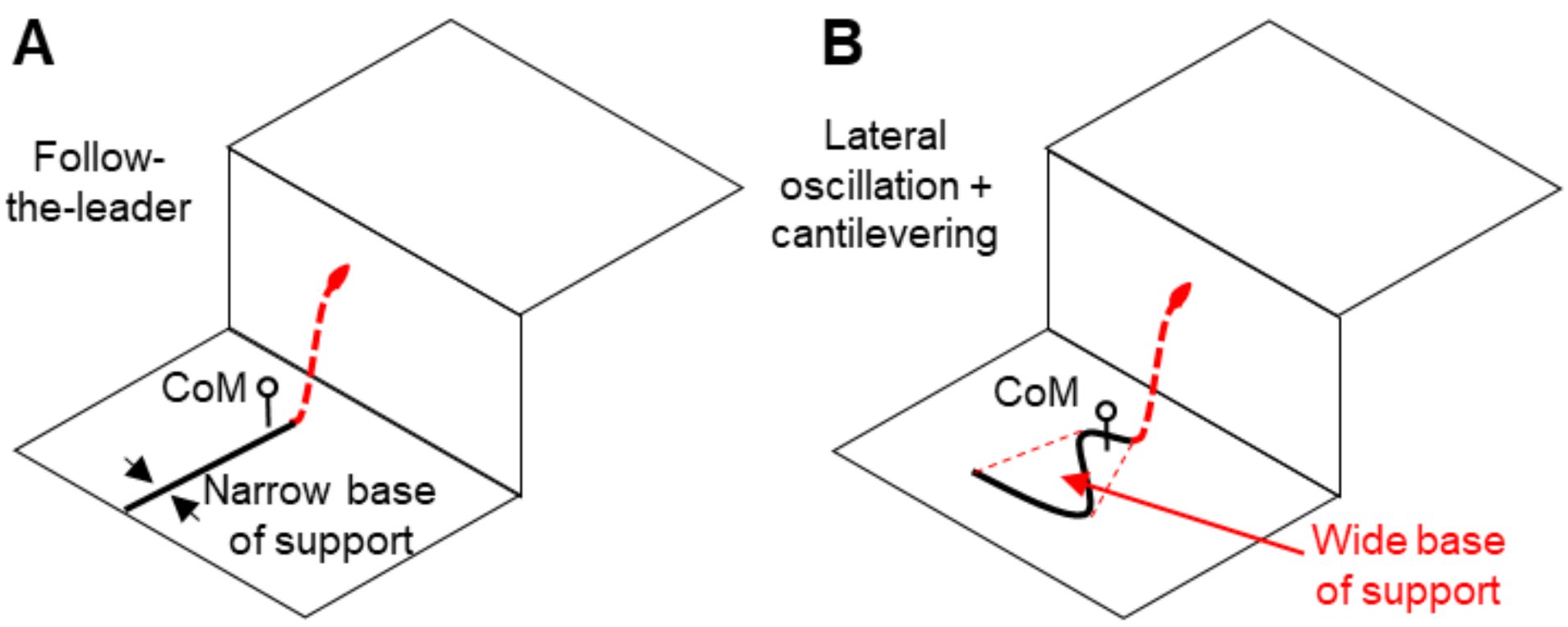

**Fig. 2. Comparison of stability of large step traversal with and without body lateral oscillation.** **(A)** Traditional follow-the-leader gait used by many previous snake robots results in a narrow base of support. **(B)** Snake's partitioned gait combining lateral oscillation and cantilevering provides a wide base of support.

### Continuous body 3-D interpolation

To analyze the snake's stability during traversal, we developed an interpolation method to reconstruct the snake's continuous body shape and motion in three dimensions (both 3-D position and 3-D orientation) (Mitchel et al. 2020). First, we represented each body segment between two adjacent markers as a quasi-static elastic rod subject to two end constraints and assumed that it only experiences elastic forces (Fig. 3A). This is obviously far from the biomechanical reality of a snake body during locomotion. The purpose of this over-simplification is to simplify the following backbone optimization. Next, we solved a hyper-redundancy problem (Chirikjian and Burdick 1995a): a continuous slender body has many configurations that can satisfy two measured end constraints (Fig. 3B). A unique configuration can be determined (dashed curve) using the method of backbone optimization (Chirikjian 2015). Finally, we applied the method of inverse kinematics to converge the interpolated backbone curve (Kim and Chirikjian 2006) towards one that satisfies the measured end constraints (Fig. 3C), thereby approximating the midline (Fig. 3A, inset). The continuous midline of the entire body was obtained by applying the above procedures piecewise to each segment between each pair of adjacent markers. See more detail of the interpolation method in (Mitchel et al. 2020).

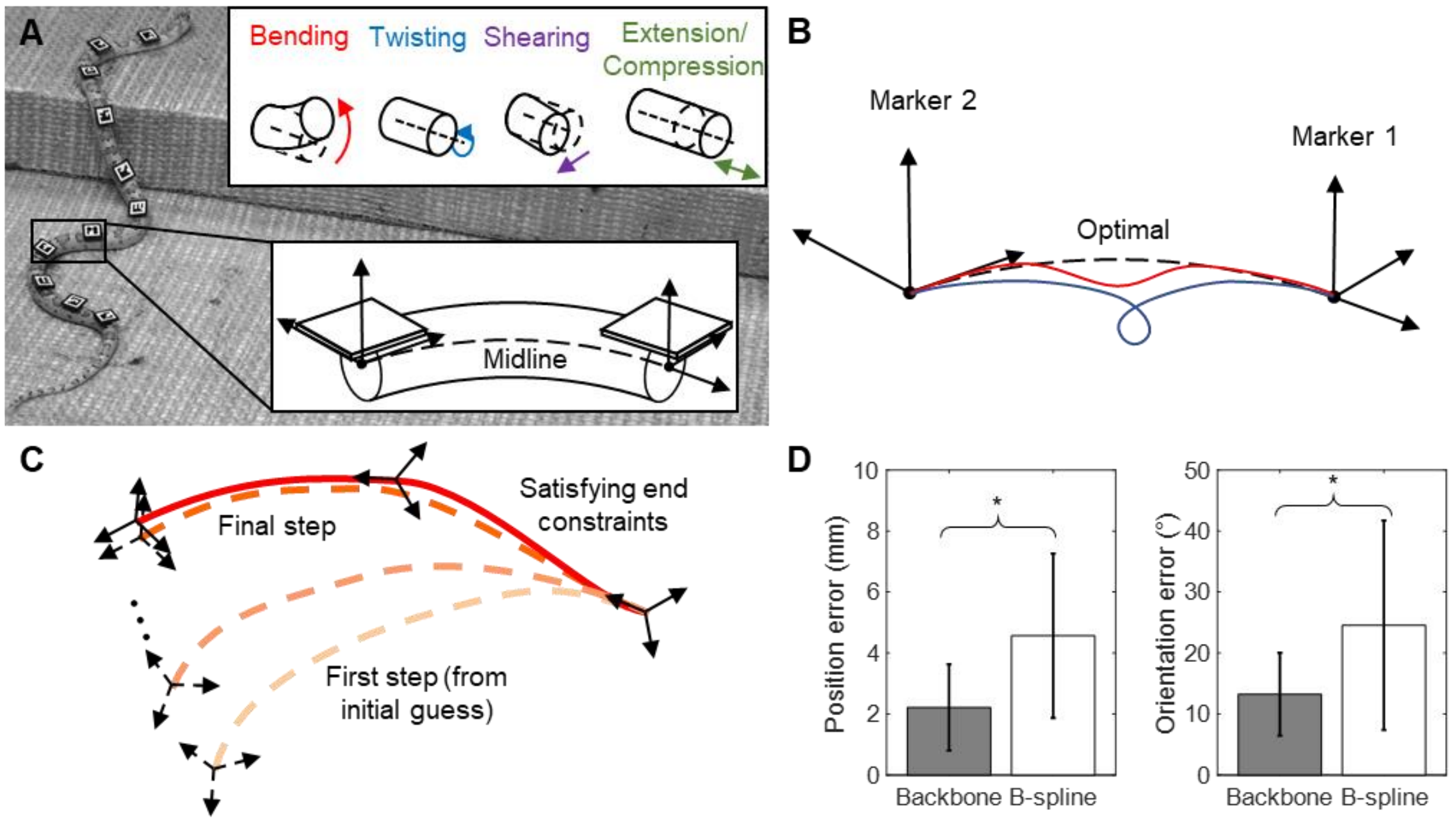

**Fig. 3. Interpolation method to reconstruct continuous body in 3-D from discrete tracked markers.** **(A)** Representing a snake body segment as an elastic rod subject to two end constraints imposed by two tracked markers. The rod can be bent, twisted, sheared, extended or compressed under external forces and torques. **(B)** Backbone optimization determines an optimal solution among many solutions. **(C)** Inverse kinematics converges backbone curve towards satisfying measured end constraints. **(D)** Comparison of interpolation accuracy between our method and B-spline (mean ± s.d.). Left: position error. Right: orientation error. Brackets and asterisks show a significant difference.

We compared interpolation accuracy of our method to B-spline, a commonly used geometric interpolation method (Fontaine et al. 2008; Padmanabhan et al. 2012; Sharpe et al. 2015; Socha et al. 2018; Schiebel et al. 2019), using the dataset of kingsnake traversing a large step obstacle. In both position and orientation, our method achieved higher accuracy with a 50% smaller error (Fig. 3D). Our interpolation method can be used to quantify continuous body in 3-D for other systems, such as snake predation (Penning and Moon 2017), root nutation (Ozkan-Aydin et al. 2019), and soft robotic arm manipulation (McMahan and Walker 2009). However, we note that this benefit comes at a cost—one needs to use markers that provide 3-D orientation as well as position information.

**Base of support analysis reveals perfect stability of kingsnakes**

Using the interpolation method, we examined the snake's static stability during large step traversal by analyzing whether its center of mass projects into the base of support formed by its body sections in contact with both horizontal surfaces below and above the step (Ting et al. 1994). Despite its large out-of-plane body bending (up to 30% SVL), in all 120 trials, the snake maintained perfect static stability regardless of step height and friction, with its center of mass projection falling into the base of support 100% of the time for all trials (see Fig. 4 for representative snapshots). This is remarkable especially for the low friction, high step, where the snake suffered large slip. The snake only occasionally braced against the vertical surface for extra support.

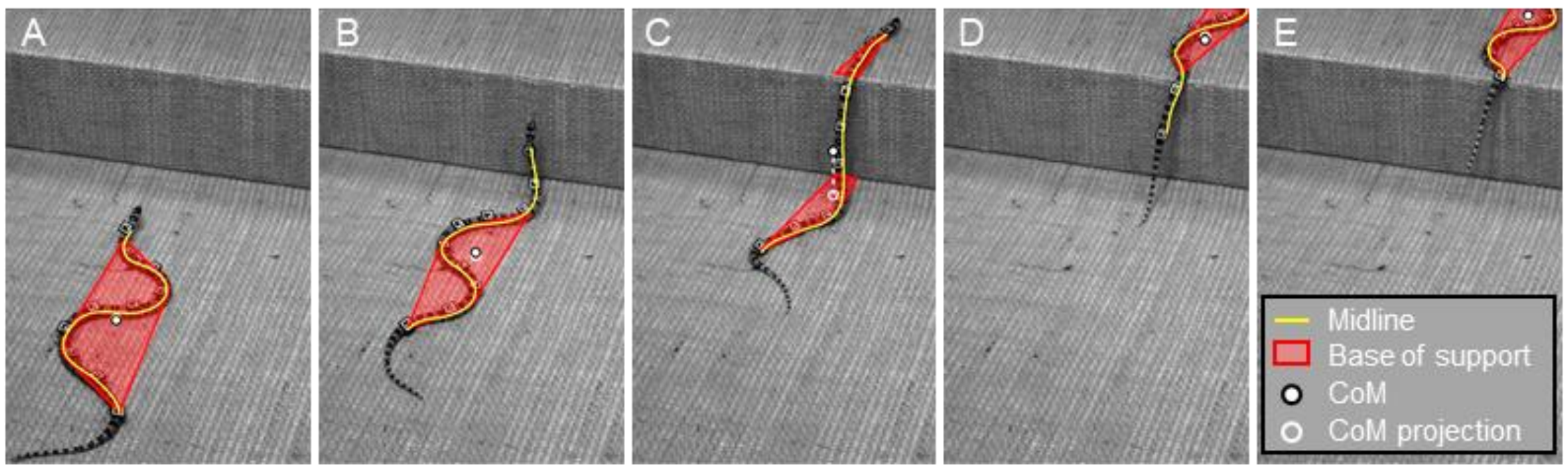

**Fig. 4. Stability analysis of snake traversing a large step.** Yellow curve is reconstructed body midline. Red curves are boundary of base of support formed by body sections in contact with horizontal surfaces, white solid circle is center of mass, and white open circle and dashed line show its projection onto horizontal surfaces.

**Snake robotic physical model reveals importance of lateral oscillation**

To test whether the lateral oscillation during cantilevering is useful for stable traversal, we developed a snake robot capable of large out-of-plane body bending (Fig. 5A) and used it as a physical model to further study the stability principles. The robot consisted 19 segments with alternating pitch and yaw joints. Each pitch segment had two one-way wheels (Fig. 5B) (Chirikjian and Burdick 1995b) to generate snake-like anisotropic friction (Fig. 5C), a feature essential for lateral undulation (Hu et al. 2009). A serpenoid traveling wave (Hirose 1993) was applied spatially to the body sections on horizontal surfaces to generate lateral oscillation, while the body section in the middle adopted a near-straight cantilevering shape. Body section division was propagated down the robot (Tanaka and Tanaka 2013) as it moved forward and upward. See (Fu and Li 2020) for more details of robot design, control, and experiments.

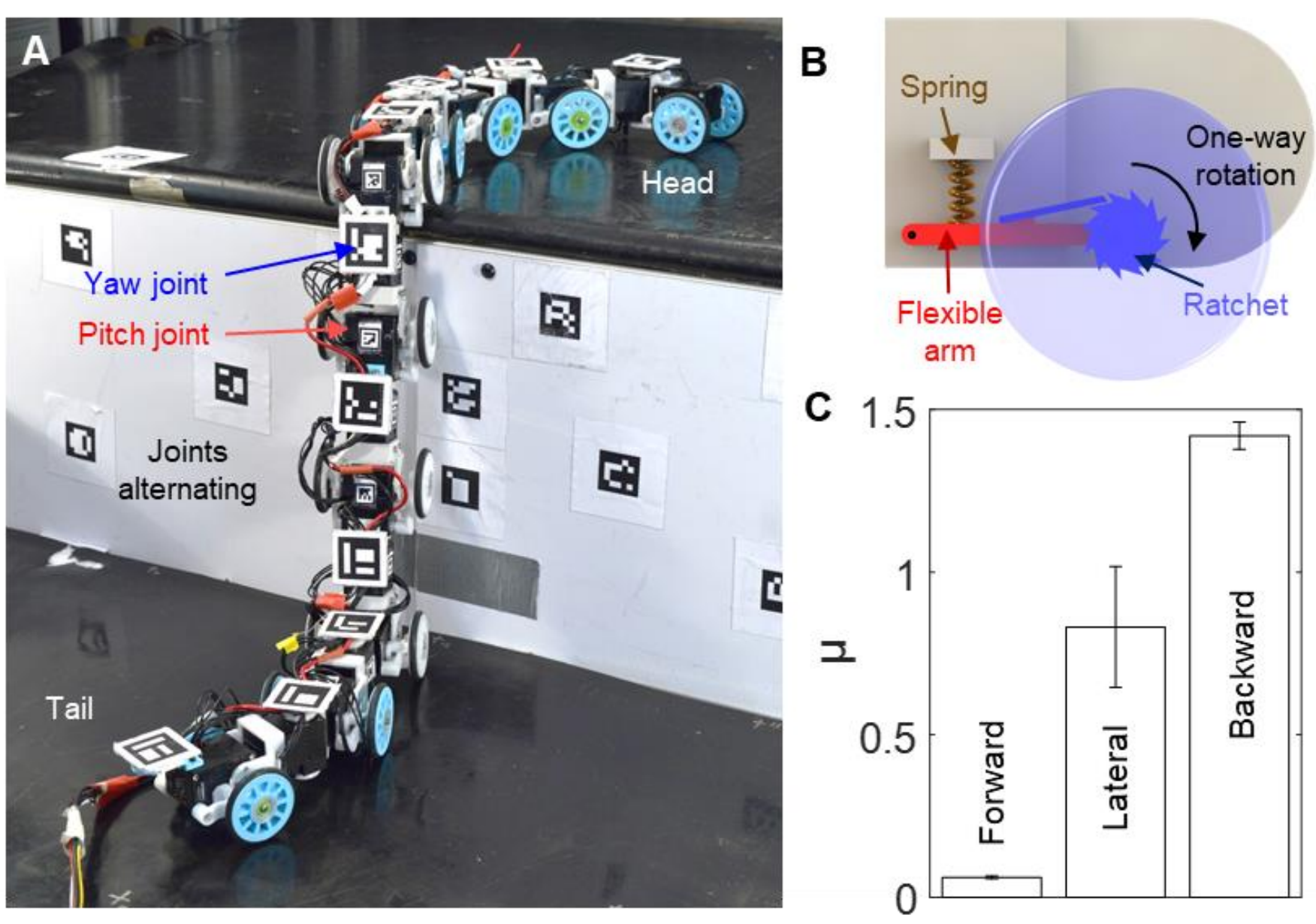

**Fig. 5. Mechanical design of snake robot. (A)** Snapshot of snake robot traversing a large step using lateral oscillation combined with cantilevering. **(B)** Design of one-way wheel for anisotropic friction and spring suspension to add body compliance to each pitch segment. Suspension can be disabled by inserting a lightweight block. **(C)** Anisotropic friction generated by one-way wheels (mean ± s.d.).

We challenged the robot to traverse a high friction step obstacle with increasingly large step height, $H$ = 31, 36, 38, and 40% of robot length $L$. Using the partitioned gait combining lateral oscillation with cantilevering from the snake, the robot rapidly traversed a step of height $H$ = 31% $L$ with probability as high as 90% (Fig. 6A, black dashed). However, traversal probability diminished quickly to 20% when step height increased to 40% $L$.

Examination of videos revealed that the majority of unsuccessful traversal was a direct result of roll failure—the robot rolled so much that it flipped over (Fig. 6C). We found that roll failure became more likely as step height increased (Fig. 6B, black dashed). This was because, as the robot devoted a longer body section to cantilevering, its oscillating sections shortened, thereby reducing the base of support. In addition, as the robot rolled, its wheels on the left and right side of the body often lost contact asymmetrically, which further worsened roll stability (Fig. 6C).

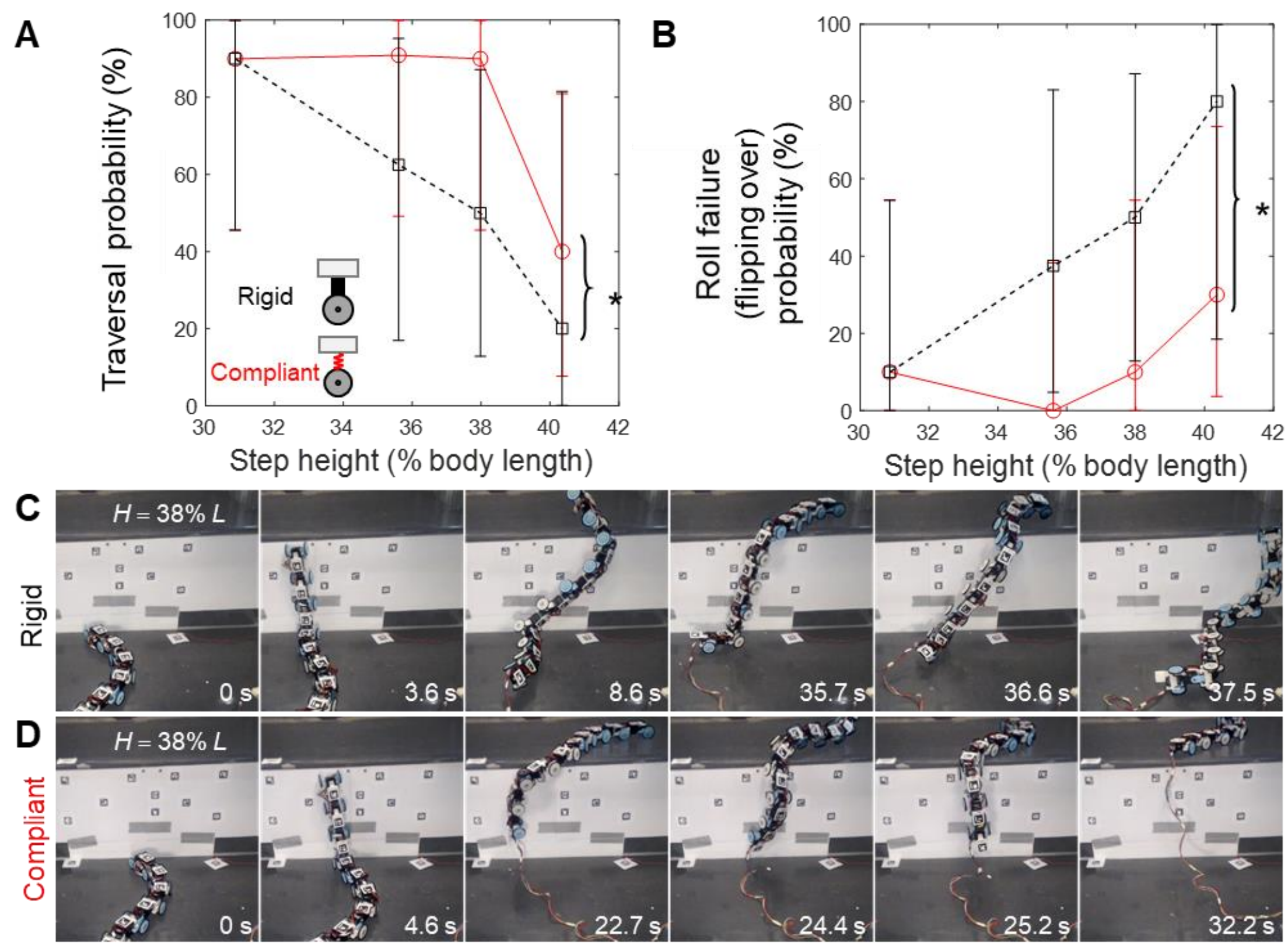

**Fig. 6. Traversal performance of robot. (A)** Traversal probability as a function of step height. **(B)** Roll failure (flipping over) probability as a function of step height. In (A) and (B), black dashed is for rigid body robot; red solid is for compliant body robot. Error bars show 95% confidence intervals. Bracket and asterisk represent a significant difference between rigid and compliant body robot. **(C, D)** Representative snapshots of robot with a rigid (C) and compliant (D) body traversing a step with a height of 38% body length.

### Animal and robot comparison suggests importance of body compliance

By contrast, even with large slipping on low friction steps, the snake never rolled as far as losing contact with the surface during traversal. The snake's better stability may be due to several reasons. First, the animal has a more continuous body (ranged between 208 and 226 pre-cloacal vertebrae (Gart et al. 2019) vs. 19 segments of the robot). In addition, the animal has a better ability to control its 3-D body bending via sensory feedback, such as flexing its head into a "hook" to gain purchase onto the top edge of the step (Fig. S1B) (Jayne and Riley 2007). Besides these, the animal's body is more compliant than the robot, with a deformable ventral side (Fig. S1A), which may result from rib motion or passive deformation of ventral tissue. This body compliance allows the animal body to deform locally to better engage the surface. Our videos that captured overall 3-D body bending did not provide sufficient spatial resolution to evaluate whether such local deformation is present in the animal. However, we could use our robot as a physical model to test the effects of body compliance.

Adding body compliance may also help increase the robot's large step traversal performance. The introduction of mechanical compliance to end effectors has proven effective in improving contact with the environment in many robotic tasks, such as grasping, polishing, and climbing (Furukawa et al. 1996; Shimoga and Goldenberg 1996; Ruotolo et al. 2019) (for a review, see (Hawkes and Cutkosky 2018)). However, despite various control compliance applied to adjust overall body shape of snake robots to adapt to the environment during locomotion (Kano and Ishiguro 2013; Travers et al. 2018), the use of mechanical compliance to better conform to surfaces was rarely considered in snake robots (but see (Togawa et al. 2000; Takemori et al. 2018b)), especially to improve stability.

**Body compliance facilitates traversal by improving contact**

To test this, we added mechanical compliance to the robot body by adding a suspension system (similar to (Togawa et al. 2000)) between each one-way wheel and its body segment (Fig. 5B). For direct comparison, the suspension system was present but disabled in rigid body robot experiments above. With the aid of body compliance, the robot was more likely to traverse higher steps than the rigid body robot (Fig. 6A, red solid vs. black dashed), succeeding with 90% probability on steps as high as 38% *L*. This was a direct result of a reduction in roll failure probability (Fig. 6B, red solid vs. black dashed) compared to the rigid body robot. By reconstructing 3-D kinematics of the robot during traversal, we verified that the compliant body robot had an improved ability to maintain contact with the step surfaces, thereby better sustaining its wide base of support for roll stability. For higher steps, the compliant body robot's traversal probability also diminished, because body segments were stuck more frequently at the top edge of the step due to smaller surface clearance resulting from compression of the suspension.

These robotic physical modeling results suggest that, besides having a more continuous body and better sensory feedback control, body compliance likely played a key role in the snake's ability to conform to the large step obstacle, maintain contact, and traverse stably.

**Our robot surpasses previous robots and approaches animal performance**

With the integration of lateral oscillation to achieve a wide base of support and one-way wheels to generate anisotropic friction for thrust, our snake robot (Fig. 7, black dashed) surpassed most previous snake robots (gray) (Yamada and Hirose 2006; Lipkin et al. 2007; Kurokawa et al. 2008; Birkenhofer 2010; Takaoka et al. 2011; Tanaka and Tanaka 2013; Tanaka et al. 2018) in traversing large step obstacles in traversal speeds (normalized to body length), approaching animal performance (blue dotted). By adding body compliance to improve surface contact, our robot further maintained high traversal probability on high steps (Fig. 6A) without suffering speed loss (Fig. 7, red solid).

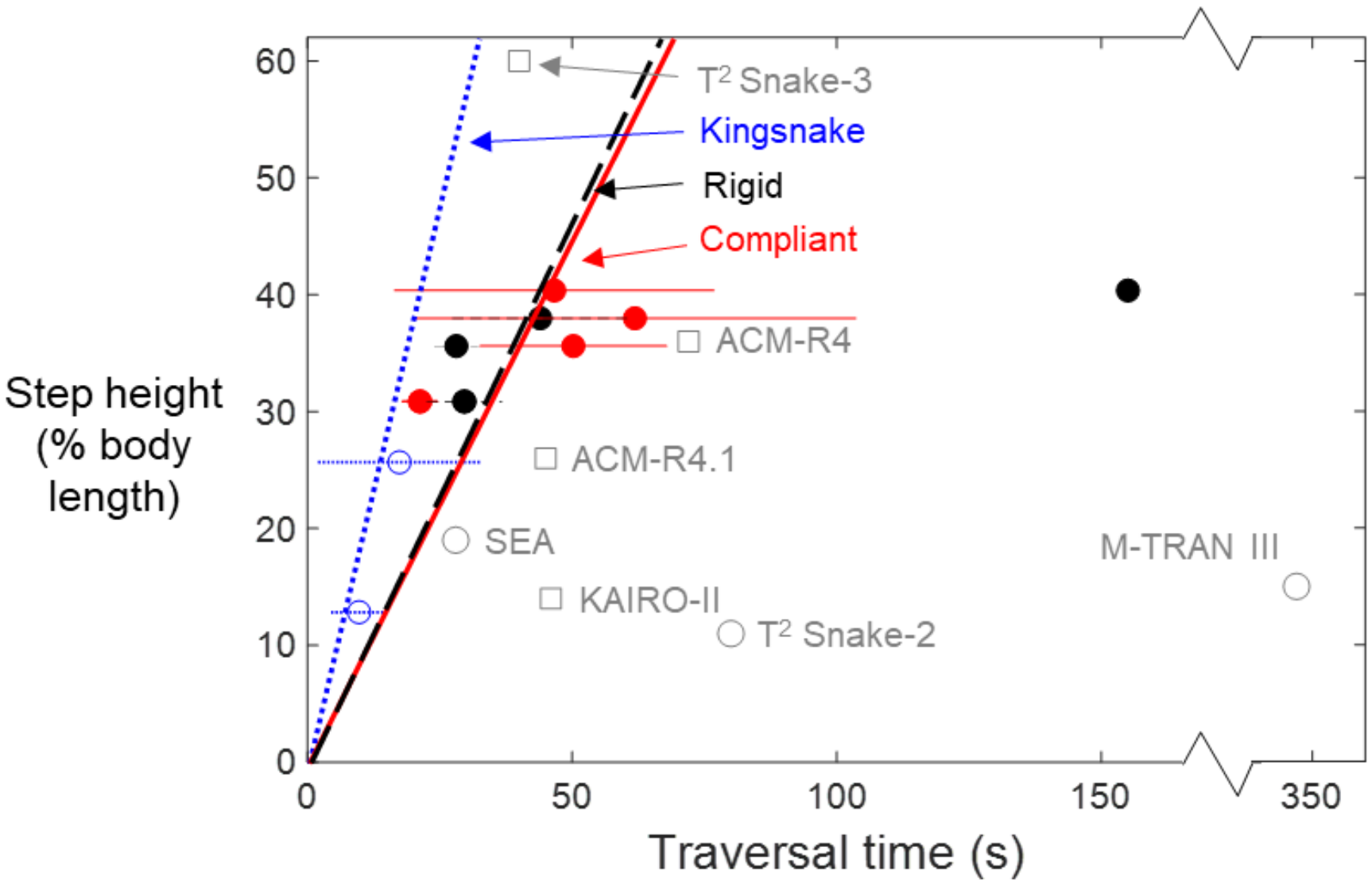

**Fig. 7. Traversable height (normalized to body length) as a function of traversal time, comparing across snake and snake robots.** Blue dotted is the variable kingsnake. Black dashed and red solid are our robot with rigid and compliant body, respectively. Gray squares and circles are previous snake robots with and without active propellers, respectively. Slopes of lines connecting data of each system to the origin show its vertical traversal speed (normalized to body length) measured over the data range. Body length is total length from head to tail for both snakes and robots.

## Summary and outlook

We discovered that the generalist kingsnake traverses large, smooth step obstacles by combining lateral oscillation and cantilevering. In addition, this partitioned gait is preserved as step height and friction vary. A base of support analysis using continuous body 3-D interpolation that we developed revealed that the animal achieves perfect static stability during traversal, even on challenging low friction, high steps. Inspired by animal observation, we developed a snake robot with lateral oscillation combined with cantilevering and snake-like anisotropic friction. The robot traversed large steps rapidly, but it suffered increasing roll failure as the step becomes higher. Adding body compliance helped the robot maintain surface contact and improve roll stability, without sacrificing traversal speed. Thanks to these, our snake robot traversed large step obstacles more rapidly (normalized to body length) than previous snake robots while maintaining high traversal probability.

The snakes still have superior traversal performance, without large body rolling (involuntarily losing surface contact) or flipping over, even on low friction steps. It will be fruitful to study how continuous body, body compliance in multiple directions, and sensory feedback control contribute to traversal of large, smooth obstacles, for which our robot provides a base platform. Finally, although we focused on a simple large step, the combination of lateral oscillation and body compliance to form a large, reliable base of support may be broadly useful to snake and snake robot moving through other large, smooth obstacles, such as stairs (Lipkin et al. 2007; Komura et al. 2015; Tanaka et al. 2018), non-parallel steps (Nakajima et al. 2018), and rubble (Sponberg and Full 2008; Takemori et al. 2018b; Whitman et al. 2018).

## Author contributions

Q.F. developed the snake robot, performed robot experiments and analysis, validated interpolation method, and drafted paper. S.W.G. performed animal experiments and analysis. T.W.M. performed animal experiments and developed interpolation method. J.S.K. and G.S.C. provided technical advice on interpolation method. C.L. supervised study and revised paper.

## Funding

This work was supported by a Burroughs Wellcome Fund Career Award at the Scientific Interface, an Arnold & Mabel Beckman Foundation Beckman Young Investigator award, The Johns Hopkins University Whiting School of Engineering start-up funds to C.L., and a Departmental Fellowship from The Johns Hopkins University Department of Mechanical Engineering to Q.F. and T.W.M.

## Acknowledgments

We thank Nansong Yi, Huidong Gao, Changxing Yan, Neil McCarter, Hongtao Wu, and Zhiyi Ren for help with apparatus development and experiments and Casey Kissel and Mitchel Stover for help with animal euthanization. We thank Ratan Othayoth, Qihan Xuan, Yuanfeng Han, Yulong Wang, Henry Astley, Bruce Jayne, David Hu, Perrin Shiebel, Dan Goldman, Jake Socha, Joe Mendelson, Bob Full, Noah Cowan, Rajat Mittal, and several anonymous reviewers for helpful comments and discussion. All animal experiments were approved by and in compliance with The Johns Hopkins University Animal Care and Use Committee (protocol RE16A223).

Special thanks to Henry Astley for helpful advice on snake care and for organizing the 2020 SICB Symposium on Long Limbless Locomotors Over Land: The Mechanics and Biology of Elongate, Limbless Vertebrate Locomotion in which this review paper was presented. We also thank the Society for Integrative & Comparative Biology, the Society for Experimental Biology, and the Company of Biologists for supporting the symposium.

**Supplementary Data**

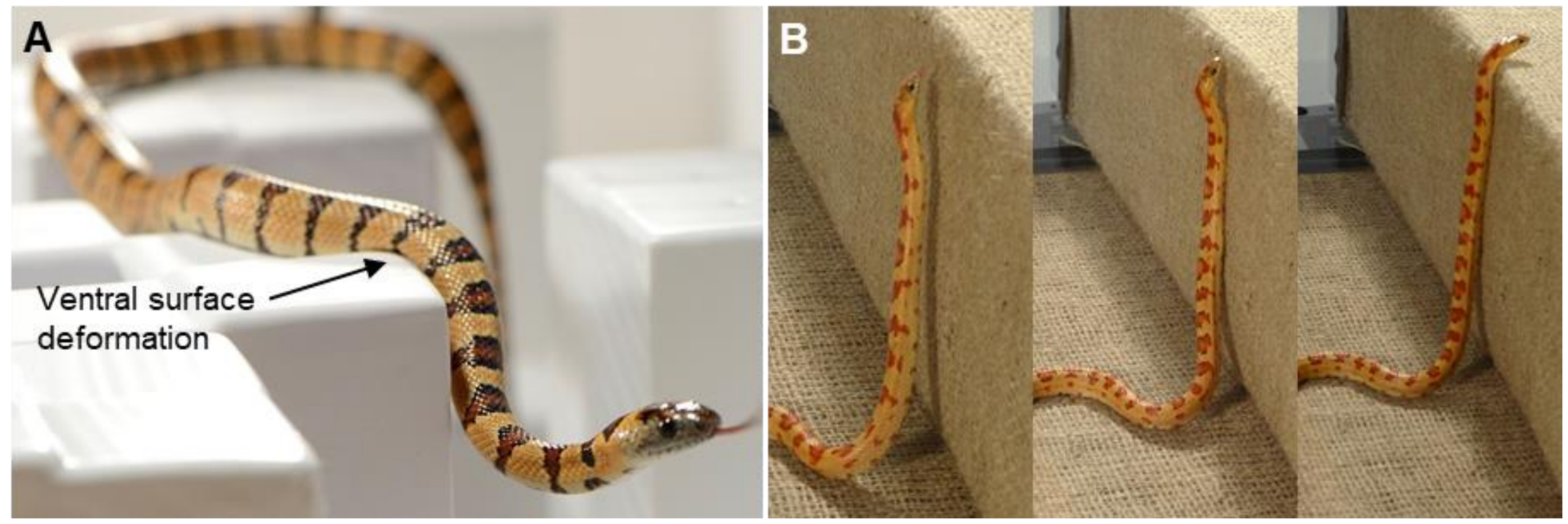

**Fig. S1. Snake body conforming to terrain using different strategies. (A)** A kingsnake engaging rubble-like terrain, showing evidence of local deformation of ventral side of the body. Photo credit: Will Kirk. **(B)** A kingsnake gaining purchase onto top edge of a step by flexing its head into a "hook".